\documentclass[conference]{IEEEtran}
\IEEEoverridecommandlockouts

\usepackage{cite}
\usepackage{amsmath,amssymb,amsfonts}
\usepackage[ruled,longend,lined]{algorithm2e}
\SetKwInOut{Input}{Input}
\SetKwInOut{Output}{Output}
\SetVlineSkip{0.5ex}

\SetCommentSty{mycommfont}

\usepackage{graphicx}
\usepackage{textcomp}
\usepackage{xcolor}
\def\BibTeX{{\rm B\kern-.05em{\sc i\kern-.025em b}\kern-.08em
    T\kern-.1667em\lower.7ex\hbox{E}\kern-.125emX}}
\begin{document}

\title{Pivot Selection for Median String Problem\\
\thanks{This work has been supported in part by CONICYT-PCHA/Doctorado Nacional/$2014-63140074$ through a Ph.D. Scholarship; Universidad Cat\'{o}lica de la Sant\'{i}sima Concepci\'{o}n through the research project DIN-01/2016; European Union's Horizon 2020 under the Marie Sk\l odowska-Curie grant agreement $690941$; Millennium Institute for Foundational Research on Data (IMFD); FONDECYT-CONICYT grant number $1170497$; and for O. Pedreira, Xunta de Galicia/FEDER-UE refs. CSI ED431G/01 and GRC: ED431C 2017/58.}
}

\author{\IEEEauthorblockN{Pedro Mirabal}
\IEEEauthorblockA{\textit{Departamento de Ingenier\'{i}a Inform\'{a}tica} \\
\textit{Unversidad Cat\'{o}lica de Temuco}\\
Chile \\
pedro.sanchez@uct.cl}

\and
\IEEEauthorblockN{Jos\'{e} Abreu}
\IEEEauthorblockA{\textit{Departamento de Ingenier\'{i}a Inform\'{a}tica} \\
\textit{Universidad Cat\'{o}lica de la Sant\'{i}sima Concepci\'{o}n}\\
Chile \\
joseabreu@ucsc.cl}

\and
\IEEEauthorblockN{Oscar Pedreira}
\IEEEauthorblockA{\textit{Laboratorio de Bases de Datos} \\
\textit{Universidade da Coru\~{n}a}\\
Spain \\
opedreira@udc.es}
}

\IEEEpubid{\makebox[\columnwidth]{978-1-7281-5613-2/19/\$31.00~\copyright 2019 IEEE \hfill} \hspace{\columnsep}\makebox[\columnwidth]{ }}

\maketitle

\begin{abstract}
The Median String Problem is W[1]-Hard under the Levenshtein distance, thus, approximation heuristics are used. Perturbation-based heuristics have been proved to be very competitive as regards the ratio approximation accuracy/convergence speed. However, the computational burden increase with the size of the set. In this paper, we explore the idea of reducing the size of the problem by selecting a subset of representative elements, i.e. pivots, that are used to compute the approximate median instead of the whole set. We aim to reduce the computation time through a reduction of the problem size while achieving similar approximation accuracy. We explain how we find those pivots and how to compute the median string from them. Results on commonly used test data suggest that our approach can reduce the computational requirements (measured in computed edit distances) by $8$\% with approximation accuracy as good as the state of the art heuristic.
\end{abstract}


\section{Introduction}
Different pattern recognition techniques such as clustering, k-Nearest Neighbors classifiers or instance reduction algorithms require prototypes to represent different patterns. Strings are widely used to represent those patterns. One particular case is contour representation using Freeman chain codes \cite{Amizam1990, Zhang2000, Rico2003a}. Also, strings are used to represent biological data such as DNA, RNA or protein sequences. The problem of finding a string that represents a set of strings has direct applications in relative compression algorithms and can be used with data such as those mentioned previously. Under Levenshtein distance, finding the median string is  W[1]-Hard for the number of strings even for binary alphabets \cite{Nicolas2005}. As an alternative to this, heuristic approaches have been proposed that iteratively refine an initial solution, applying editing operations until an approximation of the real median string is reached. However, as the size of the data sets begins to grow, the time of heuristics to find a solution to the problem increases. For that reason, focusing the search on certain elements can be fundamental to reduce the computational effort.

The Median String problem as defined in \cite{Kohonen1985} can be formalized as follows: Let $\Sigma$ be an alphabet, $\Sigma^{*}$ the set of all strings over $\Sigma$, and $\epsilon$ the empty symbol over this alphabet. For two strings $S_{i}$, $S_{j} \in \Sigma^{*}$, we denote as $E_{S_{i}}^{S_{j}}=\{e_{1}, e_{2}, ..., e_{n}\}$ the sequence of edit operations transforming $S_{i}$ into $S_{j}$. Also, let $\omega(a \rightarrow b)$ be a domain specific function that assigns a cost to an edit operation. The cost of $E_{S_{i}}^{S_{j}}$ is $\omega(E)= \sum_{e_i \in E} \omega(e_i)$ and the edit distance from $S_{i}$ to $S_{j}$ is defined as $d(S_{i}, S_{j})=argmin_{E}\{ \omega(E_{S_{i}}^{S_{j}})\}$. Given a set $S$, the string $\hat{S} \in \Sigma^{*}$ that minimizes $\sum_{S_{i} \in S} d(\hat{S},S_{i})$ is called the \textit{median string} while if it is restricted to $\hat{S} \in S$ is called the \textit{set median}. Neither the median string, nor the set median have to be unique.

\section{Related Work}\label{sec:RelatedWork}

It is a fact that we can compute the median string using the Levenshtein distance for a set of $N$ strings of length $l$ in $\mathcal{O}(l^N)$ \cite{Kruskal1983}. However, this computational cost is impractical. As an alternative, one approach consists in taking an initial string and makes successive editions over it, aiming to reduce the cumulative distance to the strings in the set. This family of algorithms is called Perturbation-based algorithms.

The most frequent choices for the initial string are the empty string and the median of the set. A ranking of possible editions allows sorting them to be applied, the way this ranking is implemented affects the computational effort. In \cite{Casacuberta1997} is described as a greedy implementation and in \cite{Kruzslicz1999} is included a tie-breaking criterion for instances where many symbols have the same goodness index.

In \cite{Kohonen1985}, the current solution is systematically changed, performing insertions, deletions, and substitutions in every position, taking as the initial string the set median. Mart\'{i}nez et.al. \cite{Hinarejos2003} use a specific order to apply operations. First, they perform substitutions in each position of the candidate solution, evaluating each possible symbol.

Every time that a new candidate solution is generated, it is necessary to compute the distance to all the strings in the set. This happens in all the cases cited above. In \cite{Fischer2000,Cardenas2004}, authors apply multiple perturbations in the same iteration. These algorithms are faster, but the quality of the approximated median that they obtain is lower.

In \cite{Bunke2002} is studied how to rank each candidate edition, for first apply the bests. \cite{Abreu2014} improved the idea of \cite{Bunke2002} and achieved more solid outcomes, increasing the convergence speed compared with \cite{Hinarejos2002} and maintaining the quality of the approximated median. One step further, in \cite{Mirabal2019}, the heuristic to select the best edit operation considerate the repercussions of each edition in all strings of the set, improving the ranking of editions. 
Another approach for the Median String Problem is as a Linear Programming problem, this is the case of \cite{Jiang2002a}, which give a lower bound and analyze the cases where the true median cannot be achieved, but do not obtain the median string. Also, \cite{Hayashida2016,Hayashida2016a}, using Integer Linear Programming, provide models for median and center string problems. In \cite{Bunke2002}, the weighted median string is presented as a generalization of this problem. In the case of, \cite{Jiang2012}, strings are embedded in a vector space, but this approach has constrictions in terms of length variation and the maximum edit distances among the set. 

In general, the whole input set $S$ is used when running the algorithms. In this work, we propose to search the approximate median, not over $S$ but in a subset $P$ such that $|P| \leq |S|$, however, ideally $|P| \ll |S|$.  The hypothesis is that computing the median only with elements in $P$ is possible to obtain a high-quality approximation to the median of $S$. This way, we expect to speed up the convergence of a reference perturbation-based heuristic while achieving similar approximation accuracy. To select $P$, we propose to apply the Spatial Selection of Sparse Pivots strategy presented in \cite{Pedreira2007} which allows us the select pivots representing the space covered by $S$.

In this work, we explored how behaves the solution in \cite{Mirabal2019} operating as it is out of the box, i.e. computing the median of $S$ using the whole set, and when is applied to $P$. Note that in both scenarios, the accuracy of the approximation is evaluated using $S$. Elements in $P$ are called pivots, in the following sections, we will describe how to select them.

\section{Reducing the problem. Search in Metric Spaces and Pivot Selection}
\label{subsec:MetricSpacesBRW}

A simple definition of a \textit{Metric Space} is as follows:

\noindent A \textit{Metric Space} is a set of elements in which the distance between them satisfies four basics rules. First, the distance between any pair of elements is not negative. Second, the distance between two elements \textit{x} and \textit{y} is \textit{0} if and only if \textit{x} and \textit{y} are the same element. Third, symmetry, the distance from \textit{x} to \textit{y} is the same that the distance from \textit{y} to \textit{x}. And fourth, triangle inequality, distance from \textit{x} to \textit{z} is always greater or equal to the sum of the distances from \textit{x} to \textit{y} and the distance from \textit{y} to \textit{z}.

\begin{enumerate}

 \item $d(x,y) \ge 0$
 
 \item $d(x,y) = 0 \leftrightarrow x = y$
 
 \item $d(x,y) = d(y,x)$
 
 \item $d(x,z) \le d(x,y) + d(x,y)$
\end{enumerate}

In this paper we use \textit{Levenshtein distance} as edit distance, which satisfies the properties of \textit{Metric Space}.

A very common problem in computer science is trying to find similar elements in a set, according to a specific function, greater than a certain threshold. The similarity function may be extremely complex in some cases, that is why to reduce the search space is a valid approach to decrease the search time.

One approach to this problem is focused on finding representatives of each region of the search space, these representatives are called pivots. When querying for the top-k similarity of a new element, the search space is reduced only to the neighborhood of the closest representative. The number of pivots varies depending on the accuracy of the similarity estimation. Normally, it is desired to have a balance between the number of pivots and the accuracy of the similarity estimation \cite{Chavez2001, Ciaccia1997}, because a large number of pivots reduces the speed of the initial search, moreover, few pivots reduce the accuracy because in this case, each pivot will represent very different elements scattered across the space.

In \cite{Pedreira2007} described the process of Sparse Pivot Selection and remarks the previous work in this field like in \cite{Baeza1994}, \cite{Baeza1997}, \cite{Mico1994} and \cite{Burkhard1973}. As we can see in the aforementioned works, the robustness of the similarity search method based on pivots depends directly on the number of pivots, the distribution between them and the distribution in the metric space. Precisely in \cite{Mico1994} the maximization of the distance between pivots is pursued, showing empirical results of the effectiveness of this method. More recently \cite{Bustos2008} presents a dynamic method of pivot selection that can modify the set of pivots while the database is growing.

\subsection{Algorithms for Pivot Selection}
\label{subsec:PivotSelection}
A good pivot selection can reduce the search space. That is why, an estimation of the maximum distance between elements is required, and for this, we use a linear algorithm. The idea of that algorithm is to calculate the maximum distance from the first element to the rest of the set, and then, do the same with the farthest element found, keeping a reference of the maximum obtained distance. This process ends when no improvement is achieved. This idea is illustrated in Algorithm \ref{algo:maxDistanceEstimation}.

\begin{algorithm}[htbp]
\LinesNumbered
\DontPrintSemicolon

\vspace{0pt}
\small
\Input{$S$}
\Output{$maxDist$}
\BlankLine
\tcc{$S$: as a list of strings.}
\BlankLine
$currentIndex = 0$\;
$formerIndex = -1$\;
$maxPossition = 0$\;
$maxDist = -\infty$\;
\BlankLine
\While{$(currentIndex \neq formerIndex)$} 
{
    \For{($i=0$  \KwTo $|S|$)} 
    {
        $dist = getDistance(S(i), S(currentIndex))$\;
        \If {$(dist > maxDist)$\;}
        {
            $maxDist = dist$\;
            $maxPossition = i$\;
        }
    }
    $formerIndex = currentIndex$\;
    $currentIndex = maxPossition$\;
}
\Return $maxDist$\;
\vspace{1ex}
\caption{maxDistanceEstimation(S) :$maxDist$}
\label{algo:maxDistanceEstimation}
\end{algorithm}

In our case, the pivot selection algorithm starts with the set median as the only element in $P$. Each element in $S$ is considered as a possible pivot if the distance to each element in $P$ is greater than a fraction of the estimated longest distance between any pair of strings in $S$. This fraction is determined by a parameter $\alpha$ which is a value between $0$ and $1$ whose optimal value is determined empirically. We also want to know how many set members are represented for each pivot. Algorithm \ref{algo:pivotSelection} shows our procedure to select pivots.

\begin{algorithm}[htbp]
 \LinesNumbered
 \DontPrintSemicolon
 \vspace{0pt}
 \small

 \Input{$S, \alpha, maxDist, setMean$ }
 \Output{$P, W$}
 \BlankLine
 \tcc{$S$: as a list of strings.}
 \tcc{$P$: list of pivots .}
 \tcc{$W$: list of respective pivot weight.}
 \BlankLine
 $P = \emptyset$\;
 $W = \emptyset$\;
 \BlankLine
 \tcc{$setMean$ is the first pivot and its weight is $1$ .}
 $P.add(setMean)$\;
 $W.add(1)$\;
 \BlankLine
 \For{($i=0$  \KwTo $|S|$)}
 {
     $posible = true$\;
     $minSpace = \infty$\;
     $pivotIndex = 0$\;
     \For{($j=0$  \KwTo $|P|$)} 
     {
         $space = getDistance(P(j), S(i))$\;
         \If{($space < maxDist*\alpha$)}
         {
             $posible = false$\;
             \If{($space < minSpace$)} 
             {
                 $minSpace = space$;
                 $pivotIndex = j$\;
             }
         }
     }
     \If{($posible$)}
     {
         $P.add(S(i))$\;
         $W.add(1)$\;
     }
     \Else
     {
        $++W(pivotIndex)$\;
     }
 }
 \Return $P, W$ \;

\vspace{1ex}
\caption{pivotSelection($S, \alpha, maxDist, setMean$) :$P,W$}
\label{algo:pivotSelection}
\end{algorithm}

To apply the pivot selection algorithm to the median string problem we propose Algorithm AppMedianStringRepercussion refering to the algorithm proposed in \cite{Mirabal2019}


\section{Experimental Results}\label{sec:exp}
We conduct experiments to compare the quality of the approximated median for $S$ obtained using pivots in $P$ respect to the reference algorithm \cite{Mirabal2019} which operates over the whole $S$. We used the average distance to the median (MAD) as a quality measure and the number of edit distance computed while the algorithms are running as a measure of the speed. 

For experimental evaluation, we work with strings that represent contours of letters using Freeman chain codes for hand-printed letters from a subset of the NIST 19 \footnote{NIST 19 supersedes NIST Special Databases 3 and 7.} special database. This codification is also used in \cite{Jain1997,Garcia-Diez2011,Rico2012,Abreu2014,Mirabal2019}. For Freeman chain codes, substitution cost between symbols is equivalent to one unit for every $45$ degrees of difference in the orientation of each symbol, for insertions and deletions the cost is always of two units as in \cite{Rico2012,Abreu2014,Mirabal2019}. We evaluated sets of size of $360$ strings for each letter of the English alphabet, thus $26$ independent samples were drawn and the average length of strings in each one computed. Attending to its average length, we binned each sample into one of the three equal-width categories short, medium and large \footnote{Using pandas.cut function}. This allows us to analyze the behavior of our approach as regards the average length of strings. For experiments, we selected some datasets from each category. For short length, we used letters P, O and I, for medium length we used R, D, B, and A, and for large length, we used W and M.

Since the size of $P$ depends on $\alpha$, we evaluate different values for this parameter, from $0.30$ to $0.02$ with a step of $0.005$. However, for each set, we report only values for $\alpha$ that lead to the most significant changes in the results.

From table \ref{tbl:P} to table \ref{tbl:W} we show the number of edit distances (Operations) required ($ \times 10^{6}$), the MAD and the size (Pivots$\%$) of $P$ as percent of the size of $S$. We also show the result of the reference algorithm running over the whole set $S$. For the maximum distance estimation and pivot selection, we count the number of distances computed for comparative reasons. We add it to the result of the median string distance count. Also, the MAD is calculated using all $S$ and not only the pivots. In bold, lowest Pivots$\%$, i.e the highest reduction, that led to a value of MAD only different to the reference in decimal order.

\vspace{1em}

\begin{table}[htbp]
\caption{Median String Dataset P.\label{tbl:P}}
 \begin{tabular}{l@{ }c@{ }c@{ }c@{ }c@{ }c}
 \vspace{0.5em}
 
$\alpha$    &0.15   &0.13   &\textbf{0.12}      &0.11   &Reference\\
\vspace{0.5em}
Pivots$\%$  &68.61  &84.72  &\textbf{90.00}     &97.22  &-\\
\vspace{0.5em}
Operations  &0.65   &0.82   &\textbf{0.79}      &0.87   &0.91\\
\vspace{0.5em}
MAD         &91.31  &90.16  &\textbf{89.92}     &89.73  &89.84\\
 \end{tabular}
\end{table}

\begin{table}[htbp]
\caption{Median String Dataset O.\label{tbl:O}}
\begin{tabular}{l@{\hspace{1em}}c@{ }c@{ }c@{ }c@{ }c}
\vspace{0.5em}
$\alpha$    &0.15   &0.12   &\textbf{0.11}   &0.10   &Reference\\
\vspace{0.5em}
Pivots$\%$  &47.78  &77.22  &\textbf{86.11}   &91.94  &-\\
\vspace{0.5em}
Operations  & 324.94 & 594.06 &\textbf{538.96} & 626.99 & 536.12\\
\vspace{0.5em}
MAD         &67.46  &63.05  &\textbf{62.50}  &62.06  &62.20\\
\end{tabular}
\end{table}

\begin{table}[htbp]
 \caption{ Median String Dataset R.\label{tbl:R}}
 \begin{tabular}{l@{ }c@{ }c@{ }c@{ }c@{ }c}
 \vspace{0.5em}
 $\alpha$   &0.17   &0.16   &\textbf{0.15}   &0.10   &Reference\\
\vspace{0.5em}
Pivots$\%$  &76.94  &81.11  &\textbf{91.11}  &99.72  &-\\
\vspace{0.5em}
Operations  &1.24   &1.55   &\textbf{1.22}   &1.66   &1.60\\
\vspace{0.5em}
MAD         &132.54 &132.19 &\textbf{131.35} &131.19 &131.40\\
 \end{tabular}
\end{table}

\begin{table}[htbp]
 \caption{Median String Dataset D.\label{tbl:D}}
 \begin{tabular}{l@{\hspace{1em}}c@{ }c@{ }c@{ }c}
 \vspace{0.5em}
$\alpha$    &0.15   &0.12   &\textbf{0.10}   &Reference\\
\vspace{0.5em}
Pivots$\%$  &57.78  &81.94  &\textbf{94.72}  &-\\
\vspace{0.5em}
Operations  &0.84   &0.69   &\textbf{0.81}   &0.73\\
\vspace{0.5em}
MAD         &102.04 &98.01  &\textbf{97.36}  &97.64\\
 \end{tabular}
\end{table}

\begin{table}[htbp]
 \caption{Median String Dataset A.\label{tbl:A}}
 \begin{tabular}{l@{\hspace{1em}}c@{ }c@{ }c@{ }c}
 \vspace{0.5em}
$\alpha$    &0.30   &\textbf{0.20}   &0.15   &Reference\\
\vspace{0.5em}
Pivots$\%$  &29.17  &\textbf{81.94}  &97.5   &-\\
\vspace{0.5em}
Operations  &0.37   &\textbf{0.81}   &0.70   &0.93\\
\vspace{0.5em}
MAD         &104.24 &\textbf{96.98}  &96.42  &96.35\\
 \end{tabular}
\end{table}

\begin{table}[htbp]
 \caption{ Median String Dataset W.\label{tbl:W}}
 \begin{tabular}{l@{ }c@{ }c@{ }c@{ }c}
 \vspace{0.5em}
$\alpha$    &0.20   &0.18   &\textbf{0.1}   &Reference\\
\vspace{0.5em}
Pivots$\%$  &61.39  &75.83  &\textbf{93.33}  &-\\
\vspace{0.5em}
Operations  &3.27   &3.84   &\textbf{3.68}   &4.04\\
\vspace{0.5em}
MAD         &218.12 &215.36 &\textbf{213.64} &214.13\\
 \end{tabular}
\end{table}

Analyzing the results we find that when Pivots$\%$ decreases the MAD increases. This can be explained since in most scenarios can be expected that the lowest is $|P|$ the lower is its ability to represent $S$. However, experiments show that the best results are achieved when the percentage of pivots is between $85\%$ and $95\%$ because the MAD is very close to the MAD for the reference result while the number of operations that we need is smaller. An interesting fact is that in some cases if Pivots$\%$ grows near to $100$, the number of operations can be even larger than the obtained without the pivot selection strategy.

In six of the nine samples \{P, I, R, B, A, W\} our approach has only a fractional difference to the reference algorithm while the number of operations was about $15\%$ lower. In the other two cases, \{D, M\}, besides the number of operations was largest, the MAD was lower, i.e the quality of the median was better than the reference.

\section{Conclusions}
In this work, we studied how to speed-up a perturbation based algorithm for the median string of a set $S$, specifically the proposed in \cite{Mirabal2019}. For this purpose, we propose to run the algorithm not over $S$ but in a subset $P$, $|P| \leq |S|$. To select this set, we applied the Spatial Selection of Sparse Pivots strategy in \cite{Pedreira2007}. We evaluated the hypothesis that computing the median in $P$ is possible to obtain a high-quality approximation to the median of $S$ but much faster.

Since Pivots$\%$ depends on $\alpha$ we test different values for this parameter. In general, as $\alpha$ increases, Pivots$\%$ and the number of operations decreases, but the MAD increases. However, if we look at the values of Pivots$\%$ for those the MAD is not worst than the reference MAD (except the fractional part) we note that our approach can reduce the number of operations in $8 \%$ as average. For example, for letter P the reference MAD is $89.84$ and the number of operations $0.91 \times 10^{6}$. For $\alpha = 0.12$, we had that Pivots$\%$ is $90.00$, the MAD is $89.92$ and the number of operations $0.79 \times 10^{6}$ which is a reduction of $13.18\%$

Results also suggest that the value $\alpha$ that can lead to a reduction of the number of operations without affecting the MAD is different in each data set. We would like to explore an approach to determine a $\alpha$ that is good enough for all data sets. 

\bibliographystyle{IEEEtran}


\end{document}